\documentclass[prl,twocolumn,showpacs,preprintnumbers,amsmath,amssymb]{revtex4}
\usepackage[dvips]{graphicx}
\usepackage{dcolumn}
\usepackage{bm}
\usepackage{times}
\usepackage{mathptmx}
\usepackage{color}

\begin{document}


\title{Tuning the Ultrafast Spin Dynamics in Carrier-Density-Controlled Ferromagnets}

\author{Masakazu Matsubara$^{1}$}
 \email{masakazu.matsubara@mat.ethz.ch}
\author{Alexander Schroer$^{2}$}
\author{Andreas Schmehl$^{3}$}
\author{Alexander Melville$^{4}$}
\author{Carsten Becher$^{1}$}
\author{Mauricio Trujillo Martinez$^{2}$}
\author{Darrell G. Schlom$^{4,5}$}
\author{Jochen Mannhart$^{6}$}
\author{Johann Kroha$^{2}$}
\author{Manfred Fiebig$^{1}$}

 \affiliation{$^{1}$Department of Materials, ETH Zurich, Wolfgang-Pauli-Strasse 10, 8093 Zurich, Switzerland}
 \affiliation{$^{2}$Physikalisches Institut, Universit\"{a}t Bonn, Nussallee 12, 53115 Bonn, Germany}
 \affiliation{$^{3}$Institut f\"{u}r Physik, Universit\"{a}t Augsburg, Augsburg 86135, Germany}
 \affiliation{$^{4}$Department of Materials Science and Engineering, Cornell University, Ithaca, New York 14853-1501, USA}
 \affiliation{$^{5}$Kavli Institute at Cornell for Nanoscale Science, Ithaca, New York 14853-1501, USA}
 \affiliation{$^{6}$Max Planck Institute for Solid State Research, Heisenbergstra{\ss}e 1, 70569 Stuttgart, Germany}

\date{\today}

\begin{abstract}
Ultrafast strengthening or quenching of the ferromagnetic order of semiconducting Eu$_{1-x}$Gd$_{x}$O was achieved by resonant photoexcitation. The modification of the magnetic order is established within 3~ps as revealed by optical second harmonic generation. A theoretical analysis shows that the response is determined by the interplay of chemically and optically generated carriers in a nonequilibrium scenario \textit{beyond} the three-temperature model. General criteria for the design of spintronics materials with tunable ultrafast spin dynamics are given.
\end{abstract}



\maketitle


The demand for an ever-increasing density and speed of manipulation in magnetic information storage triggered an intense search for ways to control the magnetic moment by means other than magnetic fields. Following the pioneering demonstration of light-induced magnetization change on the sub-ps time scale \cite{Beaurepaire_1996_PRL}, the manipulation and control of magnetic order by ultrashort laser pulses has developed into an exciting research topic \cite{Kimel_2005_Nature, Stanciu_2007_PRL, Kampfrath_2011_NatPho, Radu_2011_Nature, Kirilyuk_2010_RMP}. Most of the time-resolved investigations are focused on ultrafast \textit{demagnetization} \cite{Beaurepaire_1996_PRL, Ogasawara_2005_PRL, Koopmans_2010_NatMat, Wietstruk_2011_PRL}. This involves quenching of magnetic order and is often described by a three-temperature model (3TM), where the coupled electron, spin, and lattice subsystems are each assumed to be in equilibrium at their respective temperatures \cite{Beaurepaire_1996_PRL, Koopmans_2010_NatMat}. In special cases, however, the photoexcitation leads to the nonthermal {\it generation} and/or {\it enhancement} of magnetic order by photo-controlling the exchange interaction. Enhancement is possible if the photoexcited carriers connect neighboring magnetic sites more efficiently. For example, the ultrafast generation of ferromagnetic order in a strongly correlated manganite is based on photo-activated ``double-exchange'' interaction between Mn$^{3+}$ and Mn$^{4+}$ ions \cite{Matsubara_2007_PRL}. Transient enhancement of ferromagnetism in a diluted magnetic semiconductor is mediated by photoexcited mobile carriers connecting magnetic ions (``$p$-$d$ exchange'') \cite{Wang_2007_PRL}. These observations arouse the question if criteria can be identified that allow us to {\it enhance or attenuate} the magnetic order of a system at will.

In this Letter we show that, depending on the carrier density, ultrafast strengthening or quenching of the ferromagnetic order in Eu$_{1-x}$Gd$_{x}$O can be achieved via resonant photoexcitation. The change of the magnetic order is established within 3~ps and detected via magnetization-induced optical second harmonic generation (MSHG). The spin dynamics cannot be explained by a 3TM, but is a consequence of the photoinduced non-equilibrium carrier distribution, as the theoretical analysis shows. In contrast to systems with a high carrier concentration like transition-metal, rare-earth, or certain diluted-magnetic-semiconductor ferromagnets, the sub-ps carrier dynamics in low-doped Eu$_{1-x}$Gd$_{x}$O is far from equilibrium due to a substantially longer electron thermalization time. The resulting spin dynamics is explained by the interplay of chemically and optically generated carriers and the dynamic renormalization of the RKKY-like exchange coupling. The change from ultrafast enhancement to quenching of ferromagnetism is marked by the crossover from the predominantly semiconducting to the predominantly metallic doping range of Eu$_{1-x}$Gd$_{x}$O. Aside from the generalized insight into ultrafast spin as well as charge dynamics, a key requirement for spintronics technology, our investigation provides a guide for the selection and design of materials for ultrafast optical control of magnetic order.

\begin{figure}
\includegraphics[width=8cm,keepaspectratio,clip]{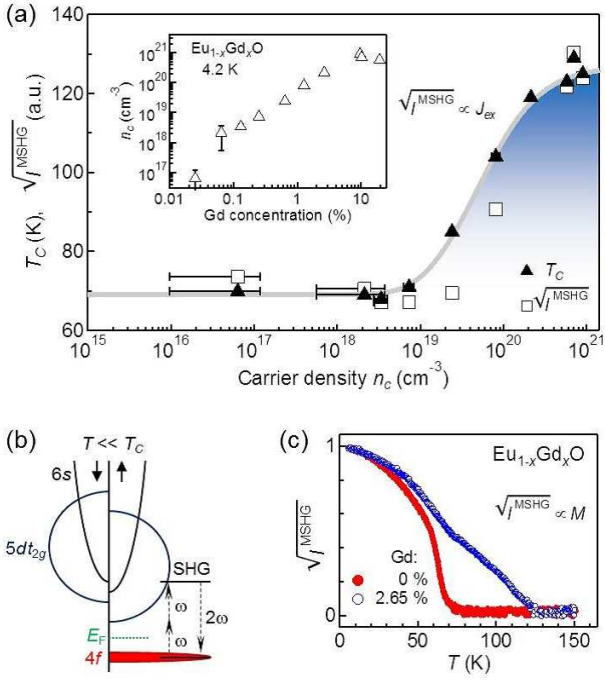}
\caption{(color online) (a) Dependence of $T_{C}$ (solid triangles) \cite{Mairoser_2010_PRL} and of the square root of the static MSHG intensity $\sqrt{I^{\rm MSHG}}$ (open squares) on the carrier density $n_{c}$ at 4.2~K in Eu$_{1-x}$Gd$_{x}$O. The inset shows the relation of the Gd concentration to $n_{c}$ \cite{Mairoser_2010_PRL}. (b) Spin-dependent electronic structure of undoped ferromagnetic EuO. The Eu $5d6s$ conduction band is split by the $4f$-$5d$ exchange interaction. The MSHG process probes the magnetic order via a two-photon transition to the $5d6s$ conduction band. A more detailed figure showing the splitting of the $4f^{6}5dt_{2g}$ sub-states can be found in Ref.~\cite{Kasuya_1972_CRCCRSSS}. (c) Temperature dependence of $\sqrt{I^{\rm MSHG}}$ for undoped and 2.65\% Gd-doped EuO.} \label{fig1}
\end{figure}

Gd-doped EuO was chosen as a candidate material because of the multitude of extreme magnetism-related properties (full spin polarization, simultaneous ferromagnetic and insulator-metal transition, colossal magnetoresistance, giant magneto-optical effects, etc.\ \cite{Schmehl_2007_NatMat, Steeneken_2002_PRL, Oliver_1972_PRB, Shapira_1973_PRB_1, Ahn_1967_IEEETM, Wang_1986_HPA, Matsubara_2010_PRB, Matsubara_2012_PRB}). Undoped EuO is a prototype Heisenberg ferromagnet. The magnetic moment of 7~$\mu_{\rm B}$ arises from the strongly localized, half-filled $4f$ shell of the Eu$^{2+}$ ions on a cubic rock salt structure. Due to the strong localization of the $4f$ orbitals, however, the direct $4f$-$4f$ exchange coupling with the nearest Eu$^{2+}$ neighbors is too weak to explain a Curie temperature as high as $T_{C} = 69$~K. Instead, the ferromagnetic order is driven by a virtual exchange mechanism \cite{Kasuya_1972_CRCCRSSS}: a virtual magnetic exciton is created by a localized $4f$ electron fluctuating into the empty Eu $5d$ state so that the magnetic exchange coupling $J_{ex}$ can be mediated via the spatially more extended $5d$ orbitals.

In Eu$_{1-x}$Gd$_{x}$O, $J_{ex}$ and $T_{C}$ are further enhanced by \textit{truly populating} the $5d$ orbitals with the extra electron introduced by replacing Eu$^{2+}$ by the otherwise magnetically equivalent Gd$^{3+}$ \cite{Sutarto_2009_PRB, Mairoser_2010_PRL}, as shown in Fig.~\ref{fig1}(a). Note that these dopant electrons do not form excitons, but rather occupy the Gd $5d$ impurity orbitals, which in the metallic phase below $T_{C}$ merge with the Eu $5d6s$ conduction band \cite{Arnold_2008_PRL} to produce a long-range RKKY contribution to $J_{ex}$. This mechanism suggests that, alternatively, an ultrafast enhancement of $J_{ex}$ may be driven by resonant optical pumping of electrons from the Eu $4f^{7}$ ground state to the $4f^{6}5dt_{2g}$ magnetic exciton state [Fig.~\ref{fig1}(b)], since this turns the virtual magnetic exciton into a real one, promoting the magnetic order.

In order to study the ultrafast spin dynamics, epitaxial Eu$_{1-x}$Gd$_{x}$O (001) films ($x = 0 - 19.5$\%) with a thickness of 35~nm were grown under adsorption-controlled conditions on two-side-polished YAlO$_{3}$ (110) single-crystal substrates by molecular beam epitaxy \cite{Mairoser_2010_PRL}. The films were protected against air by an amorphous silicon capping layer 10-20~nm in thickness and found to be free of oxygen vacancies within the resolution limit of x-ray absorption spectroscopy (XAS) \cite{Mairoser_2010_PRL}. Uniform growth without secondary phases was confirmed by x-ray diffraction, and the Gd concentration $x$ was determined by prompt-gamma activation analysis and XAS \cite{Mairoser_2010_PRL}. All Eu$_{1-x}$Gd$_{x}$O films possess an in-plane magnetic easy axis. Their physical properties were reported in Ref.~\cite{Mairoser_2010_PRL}.

\begin{figure}[b]
\includegraphics[width=\columnwidth,keepaspectratio,clip]{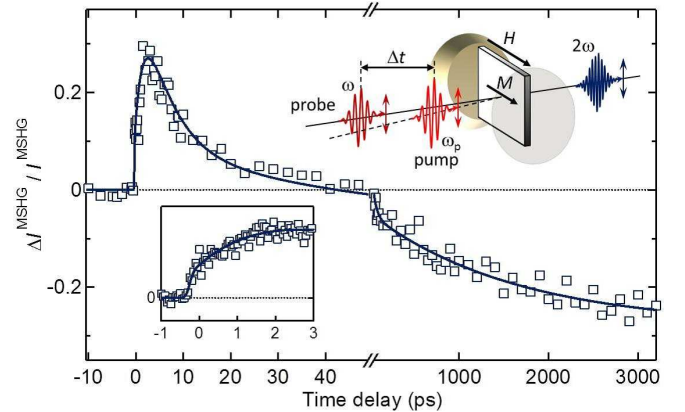}
\caption{(color online) Time evolution of the change of the MSHG intensity at 77~K for EuO doped with 2.65\% Gd. The excitation density of the pump pulse is $\sim$380~$\mu$J/cm$^{2}$. The insets show the schematic of the time-resolved MSHG spectroscopy and a magnified view of the ultrafast increase of the MSHG signal. Solid lines are guides to the eye.} \label{fig2}
\end{figure}

The experimental pump-probe setup is shown as inset in Fig.~\ref{fig2}. The output of a Ti:sapphire regenerative amplifier system at 800~nm ($\hbar\omega_{p} = 1.55$~eV) with a pulse width of 130~fs and a repetition rate of 1~kHz was divided into two beams. One was used for resonant pumping of the $4f^{7} \rightarrow 4f^{6}5dt_{2g}$ transition at 1.55~eV. The other was converted into a frequency-tunable beam by an optical parametric amplifier. This light was used for probing the time-evolution of the ferromagnetic order by MSHG. A light wave of frequency $\omega$ is incident onto the sample where it generates a polarization $P(2\omega)$ in a two-photon transition to the $5d6s$ conduction band [Fig.~\ref{fig1}(b)]. The coupling to the magnetic order occurs with high symmetry selectivity and is background-free \cite{Matsubara_2010_PRB}. The MSHG yield $I^{\rm MSHG}\propto|P(2\omega)|^2$ was measured in a normal-incidence transmission geometry under an in-plane magnetic field of 50~mT which was applied to sustain a magnetic single-domain state. The sizes of the linearly polarized probe and pump pulses were about 0.4 and 1~mm, respectively. In Appendix A we describe additional technical aspects of the MSHG measurement. In particular, we show that the MSHG signal does not display a sensitivity to the redistribution of the charge carriers by the pump pulse. We also confirmed that it originates in the bulk of the Eu$_{1-x}$Gd$_{x}$O film and not at its surface or interface. MSHG thus is an ideal probe for the dynamics of the magnetic order.

We begin by clarifying the coupling of the SHG signal to the magnetic state. (i) In Fig.~\ref{fig1}(a) we compare the dependence of $T_{C}$ and of $\sqrt{I^{\rm MSHG}}$ at 10~K where $M$ is saturated, on the Gd-related carrier density $n_c$. The two curves show the same qualitative dependence on $n_{c}$ and even quantitatively the relative difference never exceeds 25\%. According to mean-field theory, $T_{C}$ is proportional to $J_{ex}$ \cite{Arnold_2008_PRL}. With good qualitative and reasonable quantitative accuracy one may approximate the relation between $\sqrt{I^{\rm MSHG}}$ and $J_{ex}$ (resp.\ $T_{C}$) as linear. (ii) The temperature dependence of $\sqrt{I^{\rm MSHG}}$ shown in Fig.~\ref{fig1}(c) is well reproduced by that of the spontaneous magnetization $M$ \cite{Mairoser_2010_PRL}. This includes the double-dome-like dependence in the Gd-doped samples, which is related to the complementary Eu- and Gd-driven ferromagnetic ordering processes \cite{Arnold_2008_PRL} supplemented by magnetic polaron formation \cite{Tsuda_1991_book}. Thus, (i) and (ii) lead us to the phenomenological relation
\begin{eqnarray}
 I^{\rm MSHG}\propto J_{ex}^{2} M^{2}. \label{eq:IMSHG}
\end{eqnarray}
Note that, as mentioned above, the MSHG signal is insensitive to the redistribution of the charge carriers by the pump pulse. Moreover, previous experiments revealed \cite{Matsubara_2010_PRB} that at $2\hbar\omega = 2.60$~eV the interference by temperature-dependent linear absorption effects is avoided. We therefore chose it for the MSHG probe energy.

Figure~\ref{fig2} shows a typical time evolution of the photoinduced change of the MSHG intensity, normalized by the value before optical excitation, $\Delta I^{\rm MSHG}(t)/ I^{\rm MSHG}(0)$. The data were taken at 77~K on EuO doped with 2.65\% Gd. Following the photoexcitation at $t=0$ we observe a continuous {\it increase} of the MSHG intensity up to $+30$\% at $t=3$~ps. This is followed by a continuous {\it decrease} passing zero at $t=40$~ps and $-25$\% at $t=3$~ns. We conclude that the photoexcitation populates the $5d$ orbital, resulting in an enhancement of the effective $4f$-$5d$ exchange $J_{ex}$ and, by aligning the thermally fluctuating $4f$ spins, of the magnetization $M$.

The intra-atomic $4f$-$5d$ exchange energy of $\sim 0.1$~eV \cite{Kasuya_1972_CRCCRSSS} corresponds to a time of $\sim 40$~fs so that an instantaneous MSHG increase after photoexcitation might be expected in contrast to the strikingly different time of $\sim$3~ps. However, we have to take into account that the magnetic exchange coupling is mediated via the RKKY interaction. The time $\tau_{0}$ required to coherently establish its enhancement in the photoexcited region can be estimated from the time it takes the electronic correlation to spread through the crystal. The relevant distance is the RKKY wavelength, and the propagation velocity is the group velocity of the magnetic exciton (see Appendix B for numbers). The calculated buildup time $\tau_{0}$ of $\sim$10~ps is in surprisingly good agreement with the measured value of 3~ps. The ensuing decrease of MSHG intensity on the ns time scale is associated with the demagnetization by the heating of the spin system through the transfer of the optical excitation energy from the electron system to the lattice and subsequent spin-lattice relaxation. The observed slow relaxation is consistent with the empirical thermal demagnetization time \cite{Ogasawara_2005_PRL} inferred from the magnetocrystalline anisotropy constant in EuO \cite{Miyata_1967_PR}, and the decrease of the MSHG intensity is consistent with the laser-pulse heating calculated from the absorption coefficient and the heat capacity of EuO \cite{Ahn_2004_JAP}.

\begin{figure}[t]
\includegraphics[width=\columnwidth,keepaspectratio,clip]{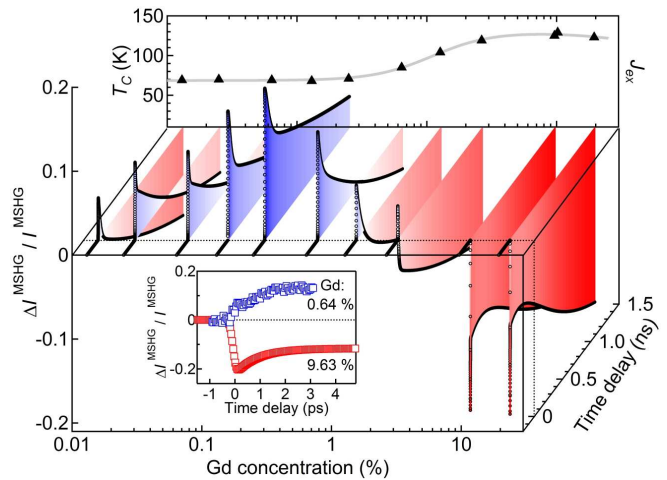}
\caption{(color online) Time evolution of the change of the MSHG intensity at 10~K for Eu$_{1-x}$Gd$_{x}$O as a function of time delay and Gd concentration. The excitation density of the pump pulse is $\sim$130~$\mu$J/cm$^{2}$ for all traces. The inset shows the magnified view of the ultrafast change of the MSHG intensity for $x=0.64$\% and 9.63\%.} \label{fig3}
\end{figure}

\begin{figure}
\includegraphics[width=\columnwidth,keepaspectratio,clip]{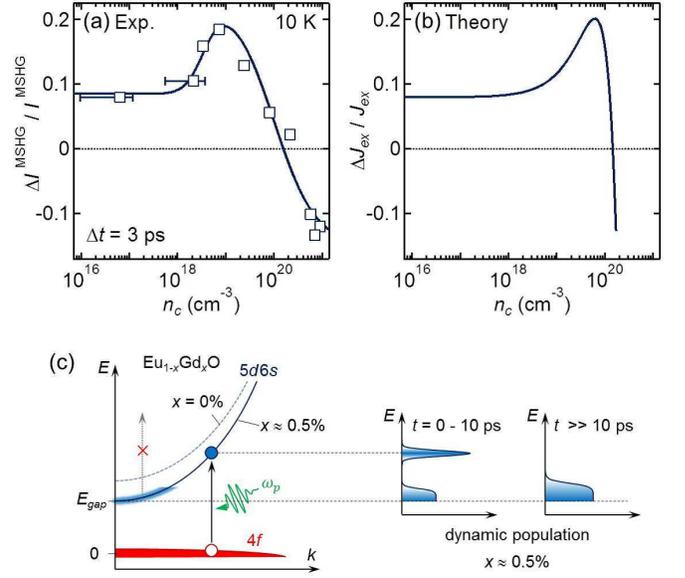}
\caption{(color online) (a) Relation between $n_{c}$ and the photoinduced change of the MSHG intensity at $t=3$~ps and at 10~K extracted from Fig.~\ref{fig3}. The line is a guide to the eye. (b) Calculated change of $J_{ex}$ due to the modification of the RKKY interaction by the photoinduced nonequilibrium population of the $5d6s$ conduction band. (c) Sketch of the photoexcitation dynamics beyond the 3TM in low-doped Eu$_{1-x}$Gd$_{x}$O. The photoinduced long-lived nonequilibrium carrier dynamics modifies the RKKY interaction, and thus $J_{ex}$. The crossed out arrow refers to photon absorption within the $5d6s$ conduction band which is kinematically forbidden. For clarity, only the majority (spin-up) carriers are shown.} \label{fig4}
\end{figure}

As the central part of the present work, we investigated the relation between the magnetic coupling dynamics and the carrier density $n_{c}$ in Eu$_{1-x}$Gd$_{x}$O. In order to simplify this investigation, the corresponding dynamics were measured at 10~K, where the magnetic moment is saturated. Hence, according to Eq.~(\ref{eq:IMSHG}), all the non-thermal changes of the MSHG intensity can then be associated with changes of $J_{ex}$. Figure~\ref{fig3} shows the temporal evolution of the normalized change of MSHG intensity induced by optical pumping for Gd concentrations between 0 and 19.5\%. A common behavior of all samples is the decreasing MSHG yield on a ns time scale. As mentioned above, this is caused by the thermal destabilization of the magnetic order mediated by the spin-lattice relaxation. In contrast, the MSHG response on the ps time scale depends strongly on the Gd concentration. At the lowest doping an ultrafast {\it increase} of the MSHG signal is observed. It strikingly confirms that our magneto-optical probe process reflects a variation of $J_{ex}$ rather than of the already saturated magnetization $M$, in contrast to the results of earlier time-resolved magneto-optical rotation measurements on undoped EuO \cite{Liu_2012_PRL}. By increasing the Gd concentration, the magnitude of this initial increase becomes progressively more pronounced and reaches a maximum near 0.25\%. Further Gd doping reduces its magnitude, and at $x\gtrsim$~5\% an ultrafast {\it decrease} of the MSHG signal is observed. In Fig.~\ref{fig4}(a) we summarize the pump-induced change of the MSHG yield as a function of $n_{c}$ at the fixed delay $t = 3$~ps. Here, $n_{c}$ has been derived for each Gd concentration $x$ from the relation displayed in the inset of Fig.~\ref{fig1}(a).

The growth of the MSHG response $\Delta I^{\rm MSHG}/I^{\rm MSHG}$ with increasing Gd concentration occurring on the ps time scale is striking. According to the widely used 3TM \cite{Beaurepaire_1996_PRL, Koopmans_2010_NatMat}, the optical pump pulse would, in addition to exciting the localized $4f$ electrons and creating the $4f^{6}5dt_{2g}$ magnetic excitons, lead to the excitation of the chemically doped carriers in the $5d6s$ conduction band and to their thermalization at the elevated temperature corresponding to the deposited energy. This would invariably {\it reduce} $J_{ex}$ since the magnetic coupling with electrons in the energetically higher band states becomes increasingly antiferromagnetic because of their increased wave number. (This was checked by explicitly calculating the RKKY interaction for the corresponding electron distribution in the conduction band: see Appendix C.) Thermalization effects would further contribute to the demagnetization.

The key to understanding the initial \textit{increase} of $\Delta I^{\rm MSHG}/I^{\rm MSHG}$ with $n_{c}$ [Fig.~\ref{fig4}(a)] is that in the low-doped Eu$_{1-x}$Gd$_{x}$O the carrier thermalization time $\tau_{c}$ is enhanced over the one in dense itinerant ferromagnets, like Ni or Gd. This is because charge thermalization occurs by electron-electron scattering whose rate is proportional to the carrier concentration. Taking the carrier thermalization time in pure Gd as $50-100$~fs \cite{Koopmans_2010_NatMat}, the $n_c$ ratio of $\approx 100$ lets us estimate $\tau_{c}$ in low-doped Eu$_{1-x}$Gd$_{x}$O to be $\tau_{c} \approx 5-10$~ps. This means that the dynamics for the first few ps is dominated by a \textit{non-equilibrium} charge carrier distribution, i.e., the 3TM is not applicable. More importantly, since electron-electron interactions are scarce in this time range, photoexcitation of the chemically doped carriers within the $5d6s$ band into higher band states is kinematically forbidden. Because of the large energy and the small momentum transferred by the photon, energy and momentum conservation cannot simultaneously be fulfilled in this two-body process. In contrast, resonant pumping of the $4f^{7} \rightarrow 4f^{6}5dt_{2g}$ transition can occur because it satisfies energy and momentum conservation [Fig~\ref{fig4}(c)]. Hence, for the first few ps after the photoexcitation, the carrier distribution is comprised of chemically doped carriers in the $5d6s$ conduction band, which remain nearly unaffected by the pump pulse, and of the fraction of the $4f$ electron population which is photo-excited into the state with $5d$ electron character. All the mobile carriers contribute to a stronger magnetic coupling between the Eu $4f$ spins.

As an additional effect, the increase in the carrier density in the conduction band strengthens its interaction with the Gd impurity band. As shown in Ref.~\cite{Arnold_2008_PRL}, this entails an energetic downshift of the conduction band toward the Gd impurity band and, hence, a reduction of the $4f$-$5d6s$ gap energy $E_{gap}$. This also affects the $4f$-$4f$ exchange as detailed in the discussion of Fig.~\ref{fig4} below.

Using the many-body renormalization theory introduced in Ref.~\cite{Arnold_2008_PRL}, we have calculated the RKKY-like coupling for the non-equilibrium electron distribution described above (see Appendix C for technical details). The result is shown in Fig.~\ref{fig4}(b). Considering the crudeness of the theoretical model, the semi-quantitative, numerical agreement with the experimental result in Fig.~\ref{fig4}(a) is compelling. Note that even without reference to the technical aspects of the calculation, the non-monotonous behavior in Figs.~\ref{fig4}(a) and \ref{fig4}(b) can be understood as follows. The density of photoexcited carriers is proportional to the spectral density at the energy $\hbar\omega_{p}-E_{gap}$ addressed by the photoexcitation. Hence, as $E_{gap}$ is reduced with increasing $n_{c}$ (see above), more states can be populated with photoexcited carriers; see Fig.~\ref{fig4}(c). This contributes to the increase of $\Delta J_{ex}(n_{c})/J_{ex}(n_{c})$ with $n_{c}$ for $n_{c} \lesssim 10^{20}$~cm$^{-3}$. For $n_{c}\gtrsim 10^{20}$~cm$^{-3}$, $\Delta J_{ex}(n_{c})/J_{ex}(n_{c})$ and $\Delta I^{\rm MSHG} / I^{\rm MSHG}$ get quenched to negative values for two reasons: (i) With further downward shift of the conduction band the energy mismatch $\hbar\omega_{p}-E_{gap}$, and with it the wave number of the photoexcited carriers, increases further, so that $J_{ex}$ acquires increasingly \textit{antiferromagnetic} RKKY-like contributions. (ii) With the increasing metallic nature the electron-electron scattering time gets reduced, leading to ultrafast thermalization and concomitant magnetization quenching, i.e., conventional demagnetization according to the 3TM begins to dominate. This is nicely reflected by the change of the buildup time from $\sim$3~ps (manifestation of RKKY interaction) to $\sim$100~fs (thermalization according to the 3TM) in the inset of Fig.~\ref{fig3}.

In summary, we have demonstrated that the ultrafast magnetic coupling dynamics can be tuned from photoinduced {\it enhancement} to photoinduced {\it quenching} of ferromagnetic order in Eu$_{1-x}$Gd$_{x}$O by controlling the carrier density. The largest enhancement of the ferromagnetic order and the crossover to quenching were observed around $n_{c} \sim 10^{19}$ and $\sim 10^{20}$~cm$^{-3}$, respectively. This behavior is explained by a nonequilibrium theory going beyond the established 3TM. Our experimental results and their theoretical modelling not only demonstrate how to control the stability of a ferromagnetic state at the ultrafast time scale, but also give a guide to the material selection and design for the dynamical optical control of magnetic order: Systems with a \textit{low density of conduction band carriers} (which is tuneable) but with a \textit{high density of magnetic moments} (unlike e.g.\ in diluted magnetic semiconductors) are in general favorable candidates.

We thank T. Stollenwerk for fruitful discussions. This work was supported by the Alexander von Humboldt Foundation, by the ETH, by the TRR 80 of the Deutsche Forschungsgemeinschaft, and by the AFOSR (Grant No. FA9550-10-1-0123).


\section{Appendix A: Probing magnetic order by second harmonic generation in E\symbol{117}O}


In this section technical details of magnetization-induced optical second harmonic generation (MSHG) in the Eu$_{1-x}$Gd$_{x}$O system are given. This includes a discussion of the coupling effects and the electronic states involved in the MSHG process. Furthermore, details on the experiment confirming that the SHG process probes the spin dynamics and is not influenced by other carrier-related nonequilibrium effects are provided.

\subsection{1. Origin of the MSHG process}

The photo-absorption and -emission constituting the SHG process can be parametrized by a multipole expansion of the coupling of the light field to the system. In the leading order of this expansion, we get electric dipole (ED) transitions and, hence, SHG is described by the relation \cite{Pershan_1963_PR},
\begin{equation}
  P_{i}(2\omega) = \epsilon_{0} \chi_{ijk} E_{j}(\omega)E_{k}(\omega).
\label{eq:ED-SHG}
\end{equation}
Here, $\vec{E}(\omega)$ represents the electric-field components of the incident light wave at $\omega$ whereas $\vec{P}(2\omega)$ represents the nonlinear electric polarization induced at $2\omega$ which acts as source of a light wave of intensity $I\propto|\vec{P}(2\omega)|^2$. The nonlinear susceptibility $\hat{\chi}$ incorporates the crystal structure, the electronic states, and the symmetry of the nonlinear material. SHG according to Eq.~(\ref{eq:ED-SHG}) is only allowed in non-centrosymmetric media. In centrosymmetric compounds like EuO, SHG becomes allowed only in higher orders of the multipole expansion, that is, to leading order, when one magnetic dipole (MD) is taken into account. In addition, ED-SHG as in Eq.~(\ref{eq:ED-SHG}) may also originate at the surface where space inversion symmetry is always broken. For centrosymmetric EuO, bulk-generated MD-SHG according to
\begin{equation}
  P_{i}(2\omega) = \epsilon_{0} ( \chi_{ijk}^{(i)} + \chi_{ijk}^{(c)} ) E_{j}(\omega)H_{k}(\omega),
\label{MD-SHG}
\end{equation}
was identified in Refs.~\cite{Matsubara_2010_PRB} and \cite{Kaminski_2009_PRL}. Here, $\hat{\chi}^{(i)}$ and $\hat{\chi}^{(c)}$ denote contributions with linear coupling to, respectively, the paramagnetic crystal or the magnetic order \cite{Fiebig_2005_JOSAB, Bennemann_1998_NOM}. In the (001)-oriented EuO films investigated by us, the non-magnetic SHG contributions according to $\hat{\chi}^{(i)}$ cannot be excited \cite{Birss_1966_book}. With the normal-incidence transmission setup sketched in the inset of Fig.~\ref{fig2}, the aforementioned surface (non-magnetic) ED-SHG contributions cannot be excited either, since in this geometry the SHG process only involves in-plane components of the light field, which are not sensitive to the inversion symmetry breaking by the surface. Thus, in our experiments, SHG should provide a background-free probe of the magnetic order. This is confirmed by the emergence of the SHG signal at $T_{C}$. Likewise, the linear dependence of the SHG intensity on the thickness of the EuO films corroborates that SHG couples to the bulk of the EuO films \cite{Matsubara_2010_PRB}.

Microscopically, MSHG in Eu$_{1-x}$Gd$_{x}$O is based on a two-photon transition from the $4f^{7}$ ground state to the $5d6s$ conduction band \cite{Matsubara_2010_PRB}. The band state at $2\hbar\omega=2.60$~eV is resonantly excited and there may be further resonance enhancement by the $4f^{6}5dt_{2g}$ exciton state near $\hbar\omega=1.30$~eV. This is reminiscent of the doubly resonant MD-SHG process observed in NiO \cite{Fiebig_2001_PRL}.

In Eu$_{1-x}$Gd$_{x}$O we identify experimentally a linear dependence of the SHG amplitude on the value of the effective exchange interaction $J_{ex}$; see main text. This is a clear indication that the two-photon absorption in Eu$_{1-x}$Gd$_x$O is assisted by an interaction process between the local spin magnetic moments. As an explanation for the linear $J_{ex}$-dependence, we thus speculate that this {\it local} interaction process provides momentum transfer to the photoexcited carriers and thus facilitates the resonant $4f$-$5d$ transition. In principle such a coupling can be expressed by expanding in powers of $J_{ex}$, and according to our experiment the leading, first-order contribution is dominant. (A detailed analysis of the two-photon excitation processes is currently in progress.)

\subsection{2. Spin dynamics versus electron dynamics}


Measurements, as in Fig.~\ref{fig1}(c), clearly demonstrate that the SHG process in Eu$_{1-x}$Gd$_{x}$O couples linearly to the magnetic order of the sample. This might, however, no longer be the case when the magnetic state of a sample is investigated in the non-equilibrium state acquired after intense photoexcitation. The redistribution of the charge carriers by the pump pulse and the ensuing relaxation dynamics may also influence the MSHG signal.

In order to scrutinize this issue we performed time-resolved linear transmission measurements on the Eu$_{1-x}$Gd$_{x}$O system. The transmittance (or reflectance) is an established probe for studying charge-carrier dynamics because it directly reproduces the population of the electronic states affected by the pump pulse \cite{Hohlfeld_1996_APB, Melnikov_2003_PRL, Satoh_2007_PRB, Matsubara_2009_PRB}. The linear transmittance at the photon energies of the fundamental and the frequency-doubled light expresses a highly complex dynamics that remains to be understood (experimental and theoretical investigations are in progress). But even without having achieved such understanding yet none of the dynamics seen in the transmittance transients is reproduced in the MSHG scan and vice versa. We can therefore conclude that the MSHG process truly probes the magnetic order whereas, within our resolution, it is insensitive to the charge carrier dynamics. With respect to the effective coupling $J_{ex}$ this means that the MSHG transient includes contributions by the virtual (undoped ground state) or the real (Gd- or photo-generated) occupation of the $5d$ orbitals.

Another carrier-related contribution to the SHG signal that might be suspected is related to the absorption of the pump light. It leads to an exponential decrease of the pump-light intensity and, hence, of the density of photoexcited carriers across the thickness of the Eu$_{1-x}$Gd$_{x}$O films. The carrier density gradient breaks the inversion symmetry and might lead to additional SHG tensor components. The presence of such contributions in the MSHG signal can, however, be excluded for three reasons. First, its temporal evolution would reflect that of the transmittance at $\omega$. Yet, such SHG contributions were not detected. Second, the absorption of the pump light is so moderate that the related carrier density changes by a factor of less than two across the films investigated by us. Third, carrier gradient contributions to SHG would occur below and above the magnetic ordering temperature. Above $T_{C}$ the SHG signal in the photoexcited samples is, however, zero at all delays.

\section{Appendix B: Time-scale of the ultrafast enhancement of $J_{ex}$}

After photoexcitation of the non-equilibrium $4f$-$5d$ exciton population, it will take a certain time $\tau_{0}$ for the magnetic RKKY correlations in the $5d6s$ conduction band and, hence, for the $J_{ex}$ enhancement to be established. As stated in the main text, $\tau_{0}$ can be estimated from the time it takes the electronic correlation to spread through the system. The relevant distance is the RKKY wavelength, and the propagation velocity is the group velocity of the magnetic exciton, which is effectively reduced with respect to the conduction electron group velocity by the ratio $m^{\ast}_{e}/m^{\ast}_h$ of the effective masses of the $5d6s$ conduction electrons ($\sim m^{\ast}_e$) and of the heavy holes in the $4f$ band ($\sim m^{\ast}_h$), which together form the magnetic exciton. This effective mass ratio was taken to be equal to the inverse ratio of the respective band widths and was extracted from literature data \cite{Arnold_2008_PRL} to be in the order of $\approx 4\times 10^{3}$. In addition, we assume a parabolic conduction bands (appropriate for our low excitation energy within the conduction band of $\Delta E = \hbar\omega_{p}-E_{gap} = 0.35$~eV). With $\tau_{0}=\pi/2 (m^{\ast}_{e}/m^{\ast}_{h})\hbar/\Delta E$ we thus obtain a buildup time $\tau_{0}$ of about 10~ps which is in good agreement with the measured value of $\sim$3~ps in Fig.~\ref{fig2}.

\section{Appendix C: Theory --- G\symbol{100}-doping dependence of $J_{ex}$ enhancement and non-equilibrium RKKY coupling}

The RKKY-like magnetic coupling and, hence, the photoinduced enhancement of $J_{ex}$ are proportional to the density of photoexcited non-thermalized conduction electrons which, in turn, is proportional to the conduction electron density of states, $N(\Delta E)$, at the final-state energy of the photoexcited electrons, $\Delta E = \hbar\omega_{p}-E_{gap}$. As earlier calculations have shown \cite{Arnold_2008_PRL}, this density of states is sensitively affected by the Gd-doping concentration $x$. Increasing $x$ leads to a downward shift of spectral weight and an enhancement of $N(\Delta E)$ for the photon energy $\hbar\omega_{p}=1.55$~eV used in our experiments. We have calculated the RKKY coupling produced by the photoinduced, non-equilibrium electron population in the conduction band, where the latter is renormalized by the presence of the Gd impurities as described above \cite{Arnold_2008_PRL}. For $x$ in the order of 1\%, this coupling is ferromagnetic and increasing with $x$, whereas for growing $x$ it acquires increasingly antiferromagnetic contributions. In order to extract the effective exchange coupling as a single number that can be compared with the experiment, first the Curie temperature $T_{C}$ was calculated for the long-range RKKY coupling using mean-field theory. The effective exchange coupling $\Delta J_{ex}$ as shown in Fig.~\ref{fig4}(b) was then determined as the effective nearest-neighbor coupling within a Heisenberg model displaying the same value of $T_C$. The interplay of the $x$-dependent increase of density of states and of the increasingly antiferromagnetic RKKY contributions leads to the non-monotonous behavior of $\Delta J_{ex}/J_{ex}$ depicted in Fig.~\ref{fig4}(b). Thermalization processes becoming more effective for larger $x$ may lead to an additional reduction of $\Delta J_{ex}$ at $x \gg 1$\%. Note that the theoretical curve of Fig.~\ref{fig4}(b) contains only a \textit{single} freely adjustable parameter. This is the unknown overlap integral of the conduction electron wave functions with the local $4f$ wave functions. This parameter determines the overall size of $\Delta J_{ex}/J_{ex}$, i.e., the scale on the vertical axis of Fig.~\ref{fig4}(b). The only other parameter of the model calculation is the ratio of the effective masses of the $5d6s$ conduction electrons and of the heavy holes in the $4f$ band, which was calculated in Appendix B and is thus not a fit parameter.




\newpage


\begin{thebibliography}{99}

\bibitem{Beaurepaire_1996_PRL}
E. Beaurepaire, J.-C. Merle, A. Daunois, and J.-Y. Bigot,
Phys. Rev. Lett. \textbf{76}, 4250 (1996).

\bibitem{Kimel_2005_Nature}
A. V. Kimel, A. Kirilyuk, P. A. Usachev {\it et al}., 
Nature \textbf{435}, 655 (2005).

\bibitem{Stanciu_2007_PRL}
C. D. Stanciu, F. Hansteen, A. V. Kimel {\it et al}., 
Phys. Rev. Lett. \textbf{99}, 047601 (2007).

\bibitem{Kampfrath_2011_NatPho}
T. Kampfrath, A. Sell, G. Klatt {\it et al}., 
Nature Photon. \textbf{5}, 31 (2011).

\bibitem{Radu_2011_Nature}
I. Radu, K. Vahaplar, C. Stamm {\it et al}., 
Nature \textbf{472}, 205 (2011).

\bibitem{Kirilyuk_2010_RMP}
A. Kirilyuk, A. V. Kimel, and T. Rasing,
Rev. Mod. Phys. \textbf{82}, 2731 (2010), and references therein.

\bibitem{Koopmans_2010_NatMat}
B. Koopmans, G. Malinowski, F. Dalla Longa {\it et al.}, 
Nature Mat. \textbf{9}, 259 (2010).

\bibitem{Wietstruk_2011_PRL}
M. Wietstruk, A. Melnikov, Ch. Stamm {\it et al.}, 
Phys. Rev. Lett. \textbf{106}, 127401 (2011).

\bibitem{Ogasawara_2005_PRL}
T. Ogasawara, K. Ohgushi, Y. Tomioka {\it et al}., 
Phys. Rev. Lett. \textbf{94}, 087202 (2005).

\bibitem{Matsubara_2007_PRL}
M. Matsubara, Y. Okimoto, T. Ogasawara {\it et al}., 
Phys. Rev. Lett. \textbf{99}, 207401 (2007).

\bibitem{Wang_2007_PRL}
J. Wang, I. Cotoros, K. M. Dani {\it et al}., 
Phys. Rev. Lett. \textbf{98}, 217401 (2007).

\bibitem{Schmehl_2007_NatMat}
A. Schmehl, V. Vaithyanathan, A. Herrnberger {\it et al}., 
Nature Mater. \textbf{6}, 882 (2007).

\bibitem{Steeneken_2002_PRL}
P. G. Steeneken, L. H. Tjeng, I. Elfimov {\it et al}., 
Phys. Rev. Lett. \textbf{88}, 047201 (2002).

\bibitem{Oliver_1972_PRB}
M. R. Oliver, J. O. Dimmock, A. L. McWhorter, and T. B. Reed,
Phys. Rev. B \textbf{5}, 1078 (1972).

\bibitem{Shapira_1973_PRB_1}
Y. Shapira, S. Foner, and T. B. Reed,
Phys. Rev. B \textbf{8}, 2299 (1973).

\bibitem{Ahn_1967_IEEETM}
K. Y. Ahn and J. C. Suits,
IEEE Trans. Magn. \textbf{MAG3}, 453 (1967).

\bibitem{Wang_1986_HPA}
H.-Y. Wang, J. Schoenes, and E. Kaldis,
Helv. Phys. Acta \textbf{59}, 102 (1986).

\bibitem{Matsubara_2010_PRB}
M. Matsubara, A. Schmehl, J. Mannhart {\it et al}., 
Phys. Rev. B \textbf{81}, 214447 (2010).

\bibitem{Matsubara_2012_PRB}
M. Matsubara, A. Schmehl, J. Mannhart {\it et al}., 
Phys. Rev. B \textbf{86}, 195127 (2012).

\bibitem{Kasuya_1972_CRCCRSSS}
T. Kasuya,
CRC Crit. Rev. Solid State Sci. \textbf{3}, 131 (1972).

\bibitem{Sutarto_2009_PRB}
R. Sutarto, S. G. Altendorf, B. Coloru {\it et al}., 
Phys. Rev. B \textbf{79}, 205318 (2009).

\bibitem{Mairoser_2010_PRL}
T. Mairoser, A. Schmehl, A. Melville {\it et al}., 
Phys. Rev. Lett. \textbf{105}, 257206 (2010).

\bibitem{Arnold_2008_PRL}
M. Arnold and J. Kroha,
Phys. Rev. Lett. \textbf{100}, 046404 (2008).

\bibitem{Tsuda_1991_book}
N. Tsuda, K. Nasu, A. Yanase, and K. Siratori, \textit{Electronic Conduction in Oxides}, Springer
Series in Solid-State Sciences Vol.\ 94, (Springer, Berlin, 1991).

\bibitem{Miyata_1967_PR}
N. Miyata and B. E. Argyle,
Phys. Rev. \textbf{157}, 448 (1967).

\bibitem{Ahn_2004_JAP}
K. Ahn, A. O. Pecharsky, K. A. Gschneidner, Jr., and V. K. Pecharsky,
J. Appl. Phys. \textbf{97}, 063901 (2004).

\bibitem{Liu_2012_PRL}
F. Liu, T. Makino, T. Yamasaki {\it et al}., 
Phys. Rev. Lett. \textbf{108}, 257401 (2012).



\bibitem{Pershan_1963_PR}
P.S. Pershan,
Phys. Rev. {\bf 130}, 919 (1963).

\bibitem{Kaminski_2009_PRL}
B. Kaminski, M. Lafrentz, R. V. Pisarev, D. R. Yakovlev, V. V. Pavlov, V. A. Lukoshkin, A. B. Henriques, G. Springholz, G. Bauer, E. Abramof, P. H. O. Rappl, and M. Bayer,
Phys. Rev. Lett. \textbf{103}, 057203 (2009).

\bibitem{Fiebig_2005_JOSAB}
M. Fiebig, V. V. Pavlov, and R. V. Pisarev,
J. Opt. Soc. Am. B \textbf{22}, 96 (2005).

\bibitem{Bennemann_1998_NOM}
\textit{Nonlinear Optics in Metals}, edited by K. H. Bennemann (Clarendon Press, Oxford, 1998).

\bibitem{Birss_1966_book}
R. R.Birss, \textit{Symmetry and Magnetism}, (North-Holland, Amsterdam, 1966).

\bibitem{Fiebig_2001_PRL}
M. Fiebig, D. Fr\"{o}hlich, Th.\ Lottermoser, V. V. Pavlov, R. V. Pisarev, and H.-J. Weber, Phys.
Rev. Lett. {\bf 87}, 137202 (2001).

\bibitem{Hohlfeld_1996_APB}
J. Hohlfeld, U. Conrad, and E. Matthias,
Appl. Phys. B: Lasers Opt. \textbf{63}, 541 (1996).

\bibitem{Melnikov_2003_PRL}
A. Melnikov, I. Radu, U. Bovensiepen, O. Krupin, K. Starke, E. Matthias, and M. Wolf,
Phys. Rev. Lett. \textbf{91}, 227403 (2003).

\bibitem{Satoh_2007_PRB}
T. Satoh, B. B. V. Aken, N. P. Duong, T. Lottermoser, and M. Fiebig,
Phys. Rev. B \textbf{75}, 155406 (2007).

\bibitem{Matsubara_2009_PRB}
M. Matsubara, Y. Kaneko, J.-P. He, H. Okamoto, and Y. Tokura,
Phys. Rev. B \textbf{79}, 140411(R) (2009).



\end{thebibliography}
\end{document}